\documentclass[journal]{IEEEtran}
\usepackage{amsmath,amssymb}
\usepackage{algorithmic}
\usepackage{algorithm}
\usepackage{array}
\usepackage[caption=false,font=normalsize,labelfont=sf,textfont=sf]{subfig}
\usepackage{textcomp}
\usepackage{stfloats}
\usepackage[hyphens]{url}
\usepackage{verbatim}
\usepackage{graphicx}
\usepackage{booktabs}
\usepackage{multirow}
\usepackage{tabularx}
\usepackage{xcolor}
\usepackage{cite}
\usepackage{float}
\usepackage{needspace}
\newcommand{\cmark}{$\checkmark$}
\newcommand{\xmark}{$\times$}
\begin{document}
\clubpenalty=10000 \widowpenalty=10000

\title{CESBench: Benchmarking Large Language Models on Cryptographic Engineering Security for IoT Devices}

\author{Wenquan~Zhou,
        An~Wang,
        Jing~Liang,
        Peien~Feng,
        Jingqi~Zhang,
        Yaoling~Ding,
        and~Liehuang~Zhu,~\IEEEmembership{Senior Member,~IEEE}
\thanks{Manuscript received [date]; revised [date]; accepted [date].}
\thanks{This work was supported in part by the National Natural Science Foundation of China under Grant 62272047 and Grant 62502035, in part by the Beijing Natural Science Foundation under Grant L244044 and Grant L251068, and in part by the State Key Laboratory of Cryptography and Digital Economy Security, Shandong University, under Grant KFZD2503. \emph{(Corresponding author: Yaoling Ding.)}}
\thanks{Wenquan Zhou, An Wang, Jing Liang, Jingqi Zhang, Yaoling Ding, and Liehuang Zhu are with the School of Cyberspace Science and Technology, Beijing Institute of Technology, Beijing 100081, China (e-mail: dyl19@bit.edu.cn).}
\thanks{Peien Feng is with the Xuteli School, Beijing Institute of Technology, Beijing 100081, China.}
\thanks{Wenquan Zhou and An Wang are also with the State Key Laboratory of Cryptography and Digital Economy Security, Shandong University, Qingdao 266237, China.}}

\markboth{IEEE Internet of Things Journal,~Vol.~XX, No.~X, Month~2026}%
{Zhou \MakeLowercase{\textit{et al.}}: CESBench: Benchmarking LLMs on Cryptographic Engineering Security for IoT Devices}

\maketitle

\begin{abstract}
For Internet of Things (IoT) devices, a secure algorithm alone is not enough: an attacker with physical access can attack the implementation directly, and its flaws are hard to fix once deployed. Large language models (LLMs) are now used to build and analyze such implementations. LLM benchmarks exist for cryptography and general cybersecurity, but none covers cryptographic engineering. In this paper, we present \textbf{CESBench}, 380 expert-written items across six sub-domains of cryptographic engineering security for IoT devices: side-channel, fault injection, implementation, countermeasures, evaluation, and integration. Four task types target different competences: 209 multiple-choice items test recall, 67 judgment items require a security verdict and its justification, 63 scenario items require an engineering diagnosis, and 41 code tasks are graded by 572 test cases. To validate the benchmark, 11 open-weight and proprietary LLMs answer every item. Multiple-choice and code responses are scored automatically, and judgment and scenario responses by an LLM judge, whose scores are checked against a second judge from another model family and human re-scoring. Composite scores range from 54.4\% to 83.6\%. The top score on each task type is 98.6\% for multiple choice, 95.1\% for code, and 88.4\% for scenario diagnosis, but only 58.8\% for judgment. Across models, 88.5\% of verdicts are correct, yet their justifications earn only 53.4\% of the rubric marks. Multiple choice is near its ceiling for the strongest models and most code tasks are solved, whereas justifying a security verdict remains the weakest competence. The benchmark, prompts, and per-item results are public.
\end{abstract}


\section{Introduction}\label{sec:intro}

\IEEEPARstart{F}{or} Internet of Things (IoT) devices, a mathematically sound algorithm is not enough. A device in the field can end up in an attacker's hands, and its implementation can then be attacked directly: side-channel analysis recovers AES keys from a microcontroller by measuring its power consumption~\cite{kocher1999, mangard2007power}, and fault injection bypasses signature verification by glitching a clock line~\cite{joye2012fault}. A flaw in hardware or boot code that surfaces after deployment is hard to fix. Implementations are therefore evaluated before they ship, under certification regimes such as ISO/IEC 19790~\cite{iso19790} for the cryptographic module and the IoT-specific scheme SESIP~\cite{sesip2020} for the platform around it, and meeting these regimes demands scarce expertise.

That expertise is exactly what teams are now trying to source from large language models (LLMs). LLMs have moved from general coding assistance into IoT systems~\cite{rivkin2025} and IoT security workflows~\cite{zeng2025}, and they can now be asked to serve both sides of device security. On the defense side, they review constant-time code, propose countermeasures, interpret leakage-assessment results, and draft certification documentation. On the attack side, they explain leakage models, write power-analysis scripts, and suggest fault-injection parameters. In cryptographic engineering these are two faces of one discipline: the knowledge a vendor needs to protect an implementation is the same knowledge an attacker uses to break it. The stakes of this delegation are high in IoT: an LLM that gives a vendor confident but wrong defensive advice bakes a flaw into silicon or boot code that ships with every unit, and an LLM that lowers the expertise needed for a side-channel or fault attack widens the pool of adversaries against every deployed unit. Yet whether LLMs are reliable in any of these roles has, to our knowledge, not been measured.

Existing benchmarks, compared in Table~\ref{tab:related}, fall short for three structural reasons. (i)~\emph{Scope.} Cryptography benchmarks target mathematical and classical cryptography, as AICrypto~\cite{aicrypto2025} and CipherBank~\cite{cipherbank2025} do, question answering on cryptography, as CryptoQA~\cite{cryptoqa2025} does, ciphertext-level cryptanalysis~\cite{maskey2025}, or recovery of algorithms from binaries, as CREBench~\cite{crebench2026} does. Hardware-security benchmarks list cryptographic and physical attacks among their categories, but test them only through code generation, as HardSecBench~\cite{hardsecbench2026} does, or through jailbreak prompts, as HarmChip~\cite{harmchip2026} does, and general security benchmarks~\cite{ctibench2024, secbench2024, cybercertbench2026} treat cryptography as at most a handful of knowledge items. IoT-oriented LLM work deploys assistants~\cite{zeng2025} or evaluates them on network-layer threat detection~\cite{tejero2025iotlogs}, not on the cryptographic module itself. None of this work covers the full chain that a device implementation must pass, from side-channel and fault attacks through countermeasures to certification. (ii)~\emph{Task type.} Multiple choice, the dominant format of these suites, mainly tests recall, and recall saturates for frontier models~\cite{hendrycks2021mmlu}. The benchmarks that go beyond multiple choice add open-ended tasks of their own, such as written proofs and capture-the-flag solutions in AICrypto, reimplementation of algorithms recovered from binaries in CREBench, and tracing vulnerability descriptions to their root causes in CTIBench. Yet each of them covers at most two of the three task types that the engineering work above calls for: judging whether a security claim holds and why, diagnosing a concrete engineering scenario, and writing code that is executed and checked. (iii)~\emph{Scoring validity.} LLM judges carry known biases~\cite{zheng2023judge}, yet the benchmarks that use them rarely report quantitative agreement between the judge and a second judge or human raters.

\begin{table}[hbpt]
\centering
\caption{Coverage of the Three Gaps by Existing LLM Benchmarks}
\label{tab:related}
\footnotesize
\setlength{\tabcolsep}{2.6pt}
\begin{tabular}{@{}lcccccc@{}}
\toprule
 & \multicolumn{2}{c}{Scope} & \multicolumn{3}{c}{Task type} & Scoring \\
\cmidrule(l{2pt}r{2pt}){2-3}\cmidrule(l{2pt}r{2pt}){4-6}\cmidrule(l{2pt}r{2pt}){7-7}
Benchmark & Crypto & Physical & Judgment & Scenario & Code & Validated \\
\midrule
AICrypto~\cite{aicrypto2025} & \cmark & \xmark & \cmark & \xmark & \cmark & \cmark \\
CREBench~\cite{crebench2026} & \cmark & \xmark & \xmark & \cmark & \cmark & \xmark \\
CryptoQA~\cite{cryptoqa2025} & \cmark & \xmark & \xmark & \xmark & \xmark & \xmark \\
HardSecBench~\cite{hardsecbench2026} & \cmark & \cmark & \xmark & \xmark & \cmark & \xmark \\
HarmChip~\cite{harmchip2026} & \cmark & \cmark & \xmark & \xmark & \xmark & \xmark \\
CTIBench~\cite{ctibench2024} & \xmark & \xmark & \xmark & \cmark & \xmark & \xmark \\
\textbf{CESBench (ours)} & \cmark & \cmark & \cmark & \cmark & \cmark & \cmark \\
\bottomrule
\end{tabular}
\end{table}

\begin{figure}[!t]
	\centering
	\includegraphics[width=\columnwidth]{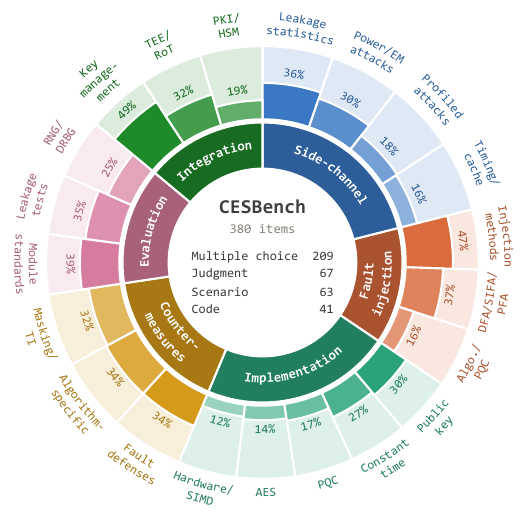}
	\caption{Composition of CESBench.}
	\label{fig:arch}
\end{figure}

To fill these gaps, we present CESBench, a benchmark of 380 expert-written items on the cryptographic engineering layer of IoT device security, and use it to measure 11 current LLMs. Fig.~\ref{fig:arch} shows the composition of CESBench: the inner ring divides the 380 items into six sub-domains, the outer ring splits each sub-domain into its topics and shows the share of items each topic holds, and the centre lists the four task types.

Our contributions are threefold:
\begin{enumerate}
    \item \textbf{A benchmark of cryptographic engineering security for IoT devices.} We build CESBench,\footnote{\url{https://github.com/wenquan222/CES}} 380 expert-written items across six sub-domains, from side-channel and fault attacks to certification and module integration, each written from the literature or the standards and reviewed by a second expert. The items cover four task types: 209 multiple-choice items, 67 judgment items, 63 scenario items, and 41 code tasks graded by 572 test cases. We also label each item by how strongly its answer depends on IoT device conditions.
    \item \textbf{A scoring protocol that separates conclusions from reasoning.} We score multiple-choice answers and code automatically, and judgment and scenario answers with an LLM judge from a separate model family. On judgment items a wrong verdict scores zero and a correct one earns marks only for its justification, so weak reasoning behind a right verdict still shows in the score. We validate the judge by re-judging, by a second judge from another family, and by blind human re-scoring.
    \item \textbf{An evaluation of 11 LLMs that locates their weakness in justification.} We evaluate 11 open-weight and proprietary LLMs, whose overall scores range from 54.4\% to 83.6\% and whose best scores reach 98.6\% on multiple choice, 95.1\% on code, and 88.4\% on scenarios, but only 58.8\% on judgment. We find that the models label a claim true or false correctly 88.5\% of the time, yet the justifications they give for these correct labels score only 53.4\%. As multiple choice no longer separates the strongest models, we argue that benchmarks should also test whether models can justify their conclusions.
\end{enumerate}

\section{Related Work}\label{sec:related}

Cryptographic engineering security is the discipline of making cryptography survive on a physical device, from leakage and side-channel and fault attacks to countermeasures, certification, and module integration. We review LLM benchmarks in three neighboring fields by what they cover, which task types they use, and how they score responses.

\subsection{LLM Benchmarks in Cryptography}

LLM benchmarks in cryptography test algorithms, ciphers, schemes, and binaries. AICrypto~\cite{aicrypto2025} combines multiple-choice questions, capture-the-flag challenges, and proofs on mathematical and classical cryptography, CryptoQA~\cite{cryptoqa2025} builds a large question-answer set from textbooks and papers, and CipherBank~\cite{cipherbank2025} and Maskey \emph{et al.}~\cite{maskey2025} test the decryption and cryptanalysis of ciphertexts. CrypFormBench~\cite{crypformbench2026} and CryptanalysisBench~\cite{cryptanalysisbench2026} target formal analysis of schemes and attacks on modern primitives, respectively, and CREBench~\cite{crebench2026} asks models to reverse-engineer cryptographic routines from binaries. AICrypto and CREBench check flags, keys, and code automatically, CryptoQA scores free text with overlap metrics and an LLM judge compared only with other judges, and AICrypto validates its proof judge against human experts. The implementation running on a device, with its leakage, faults, and countermeasures, remains open.

\subsection{LLM Benchmarks for Hardware Security}

Hardware-security benchmarks evaluate generated hardware code and responses to attack requests. HardSecBench~\cite{hardsecbench2026} lists cryptographic, power and clock, and physical-access weaknesses among its categories and grades generated hardware code in simulation. HarmChip~\cite{harmchip2026} includes side-channel and fault-injection prompts in a jailbreak benchmark, where a single LLM judge labels each response as compliant or refused. Physical attacks thus enter these benchmarks as code weaknesses or as requests to refuse, and the reasoning behind an implementation's resistance to them remains open.

\subsection{LLM Benchmarks in Cybersecurity and IoT}

Cybersecurity benchmarks cover threat intelligence and general security knowledge. CTIBench~\cite{ctibench2024} tests CVE root-cause mapping, severity prediction, and attacker attribution, SecEval~\cite{seceval2024} provides 2{,}126 multiple-choice questions with cryptography in less than 1\% of them, SecBench~\cite{secbench2024} pairs 44{,}823 multiple-choice questions with short-answer questions graded by an LLM agent, and CyberCertBench~\cite{cybercertbench2026} tests recall of certification exam content. Multiple choice dominates these suites, and cryptography appears in them as a small share of knowledge items.

IoT-oriented LLM work evaluates network traffic, device software, and smart-home control. Tejero-Fern\'andez and S\'anchez-Maci\'an~\cite{tejero2025iotlogs} and IDS-Agent~\cite{idsagent2024} detect threats in IoT logs and traffic, Rondanini \emph{et al.}~\cite{rondanini2026edge} detect malware with lightweight LLMs on edge devices, Abtahi and Azim~\cite{abtahi2025firmware} patch LLM-generated embedded firmware against common software weaknesses, and Zeng \emph{et al.}~\cite{zeng2025} and Rivkin \emph{et al.}~\cite{rivkin2025} build a security assistant and a smart-home agent, respectively. The cryptographic module inside the device remains open as an object of evaluation.

\section{Benchmark Design}\label{sec:design}

CESBench consists of 380 expert-written items on cryptographic engineering security, organized by six sub-domains and four task types: multiple choice, judgment, scenario, and code. Fig.~\ref{fig:arch} draws this composition and Table~\ref{tab:distribution} gives the item counts per sub-domain and task type. Each item also carries a label for how strongly its answer depends on IoT device conditions. Multiple-choice answers and code are scored automatically. Judgment and scenario answers are scored by an LLM judge that is validated against a second judge and a human rater.

\begin{table}[ht]
\centering
\caption{Question Distribution of CESBench}
\label{tab:distribution}
\footnotesize
\setlength{\tabcolsep}{4pt}
\begin{tabular}{@{}lccccc@{}}
\toprule
Sub-domain & Multiple choice & Judgment & Scenario & Code & Total \\
\midrule
D1 Side-channel & 42 & 17 & 15 & 6 & 80 \\
D2 Fault injection & 30 & 8 & 8 & 5 & 51 \\
D3 Implementation & 44 & 15 & 14 & 10 & 83 \\
D4 Countermeasures & 33 & 11 & 10 & 8 & 62 \\
D5 Evaluation & 30 & 8 & 8 & 5 & 51 \\
D6 Integration & 30 & 8 & 8 & 7 & 53 \\
\midrule
Total & 209 & 67 & 63 & 41 & 380 \\
\bottomrule
\end{tabular}
\end{table}

\subsection{Domain Taxonomy}\label{sec:taxonomy}

IoT security standards set out what a device must achieve through cryptographic engineering. The device baselines ETSI~EN~303~645~\cite{etsi303645}, ISO/IEC~27402~\cite{iso27402}, and NIST~IR~8259A~\cite{nistir8259a} require software integrity, secure update, protected key storage, and device identification, and the certification schemes SESIP~\cite{sesip2020} and PSA Certified~\cite{psagoals, psal3} add cryptographic functions, resistance to side-channel and perturbation attacks, and evaluation at defined assurance levels. CESBench divides this ground into six sub-domains, side-channel, fault injection, implementation, countermeasures, evaluation, and integration, whose boundaries follow the aims and scope of the \emph{Journal of Cryptographic Engineering} and the module certification standards ISO/IEC 17825~\cite{iso17825}, ISO/IEC 19790~\cite{iso19790}, and \mbox{FIPS 140-3}~\cite{fips1403}. Table~\ref{tab:iotmap} maps each sub-domain to the IoT requirements it serves.

\begin{table}[!htb]
\centering
\caption{Requirements in IoT Security Standards That Each Sub-Domain Addresses}
\label{tab:iotmap}
\footnotesize
\setlength{\tabcolsep}{4pt}
\renewcommand{\tabularxcolumn}[1]{m{#1}}
\begin{tabularx}{\columnwidth}{@{}>{\centering\arraybackslash}m{0.17\columnwidth}X@{}}
\toprule
Sub-domain & Requirement (source clauses) \\
\midrule
D1 & Keep secrets from leaking through timing, power, and electromagnetic emissions (SESIP 3.4.2; PSA Certified Level 3). \\ \addlinespace[1.8pt]
D2 & Counter perturbation and tampering by a physical attacker (SESIP 3.4.2; PSA Certified Level 3). \\ \addlinespace[1.8pt]
D3 & Provide cryptographic operations, key generation, and random numbers (SESIP 3.5; NIST IR 8259A Data Protection). \\ \addlinespace[1.8pt]
D4 & Detect or prevent physical attacks, and erase residual secrets (SESIP 3.4.2 and 3.6.4). \\ \addlinespace[1.8pt]
D5 & Be evaluated at a defined assurance level (SESIP1--SESIP5; PSA Certified levels). \\ \addlinespace[1.8pt]
D6 & Protect software integrity, update securely, store keys, and identify the device (ETSI EN 303 645 5.3, 5.4, and 5.7; ISO/IEC 27402 5.2.7; NIST IR 8259A). \\
\bottomrule
\end{tabularx}
\end{table}

\begin{figure*}[!t]
\centering
\includegraphics[width=\textwidth]{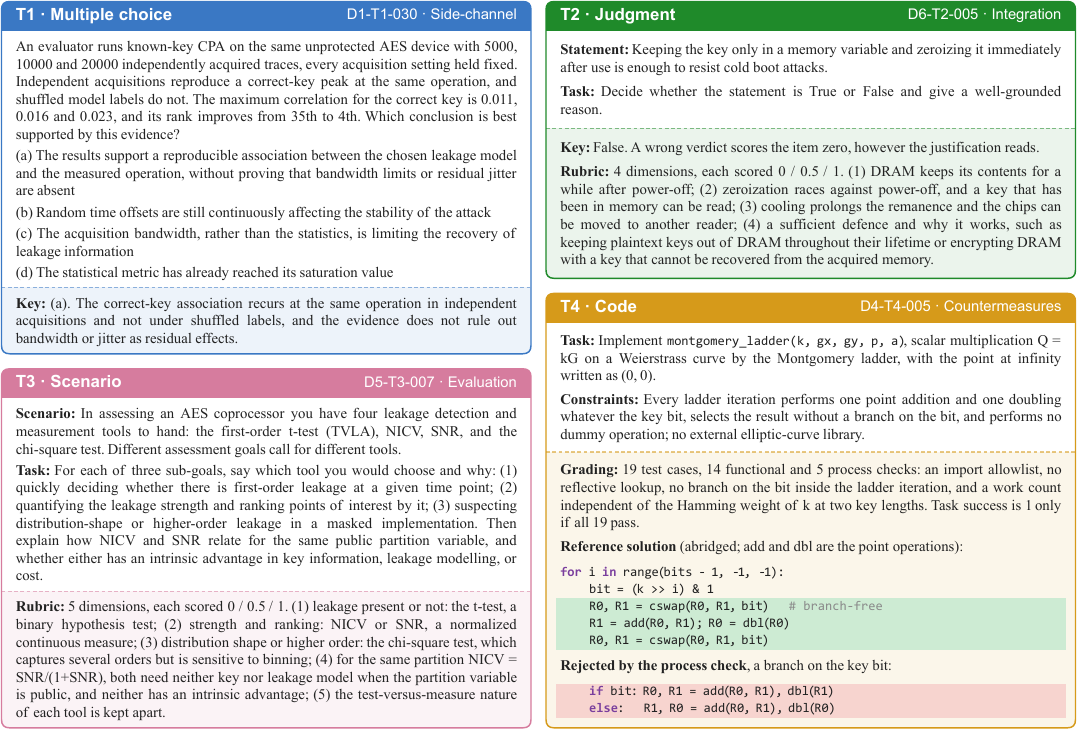}
\caption{One item per task type.}
\label{fig:examples}
\end{figure*}

\paragraph*{D1 Side-channel} The 80 items ask a model to recover or assess secrets from the physical leakage of a running implementation. The largest group, 29 items, covers leakage models and the statistics behind them, such as Hamming weight and distance models, point-of-interest selection, and signal-to-noise ratio. A further 24 items cover power and electromagnetic attacks, from simple and differential power analysis~\cite{kocher1999} and correlation power analysis~\cite{brier2004} to diagnosing why an attack fails, 14 items cover profiled, template, and deep-learning attacks~\cite{benadjila2020}, and 13 items cover timing, cache, and remote side channels~\cite{kocher1996}.

\paragraph*{D2 Fault injection} The 51 items ask a model to reason about attacks that corrupt a computation to expose secrets. The largest group, 24 items, covers injection methods and fault models, such as choosing among clock or voltage glitching and electromagnetic or laser injection, identifying the fault model, fault sensitivity analysis, remote fault injection, and one task on fault-resistant control flow. A further 19 items cover classical attacks, namely differential fault analysis (DFA) on AES~\cite{piret2003} and DES, the Bellcore attack on RSA with the Chinese remainder theorem (RSA-CRT), statistical ineffective fault attacks (SIFA)~\cite{sifa2018}, and persistent fault analysis (PFA), and 8 items cover fault attacks on specific algorithms and on post-quantum schemes.

\paragraph*{D3 Implementation} The 83 items ask a model to reason about implementing cryptographic algorithms so that they run correctly and without exploitable behavior on the target platform. The largest group, 25 items, covers public-key implementation, including Montgomery arithmetic, 22 items cover constant-time programming and the implementation attack surface, 14 items cover post-quantum cryptography (PQC) schemes such as CRYSTALS-Kyber and Dilithium, 12 items cover AES, and 10 items cover hardware implementation, bit-slicing, and vector-instruction (SIMD) acceleration.

\paragraph*{D4 Countermeasures} The 62 items ask a model to protect an implementation against the attacks of D1 and D2. Fault-injection countermeasures account for 21 items, namely error detection, infective computation, redundancy, fault sensors, and defenses against SIFA and PFA. Another 21 items cover hiding, countermeasures specific to public-key, post-quantum, and SM2/SM3/SM4 implementations, layered countermeasure architectures, and one task on erasing keys from memory. The remaining 20 items cover Boolean and arithmetic masking~\cite{isw2003}, masked AES, and threshold implementations (TI)~\cite{ti}.

\paragraph*{D5 Evaluation} The 51 items ask a model to measure whether an implementation resists attack and to certify it against a standard. Module standards and security levels account for 20 items, covering FIPS 140-3 and ISO/IEC 19790 levels, ISO/IEC 17825 procedures, Common Criteria with the Joint Interpretation Library, and GM/T 0008 levels for security chips. Leakage assessment accounts for 18 items, covering test vector leakage assessment~\cite{tvla2011, schneider2015}, leakage detection statistics, and the scope of a test platform, and 13 items cover random number generators, from statistical testing and entropy assessment to deterministic random bit generator (DRBG) constructions.

\paragraph*{D6 Integration} The 53 items ask a model to build a cryptographic module and fit it into a device and its key-management system. Key management accounts for 26 items, covering key derivation and wrapping, key hierarchies and distribution, the key lifecycle, secret sharing and backup, and key security design. A further 17 items cover trusted execution environments such as TrustZone and SGX, the root of trust and measurement, secure boot, firmware update with anti-rollback, and tamper response, and 10 items cover public-key infrastructure (PKI), hardware security modules (HSM), and compliance scope.

\subsection{Task Design}\label{sec:tasks}

CESBench uses four task types, each placing a different demand on the model. Multiple choice tests recall of the facts that cryptographic engineering rests on, judgment tests whether the model can decide if a security claim holds and explain why, scenario tests whether it can diagnose a concrete engineering situation, and code tests whether it can write an implementation that runs correctly. Fig.~\ref{fig:examples} shows one item of each type. In each card the white area is what the model receives, and the tinted strip is what the grader holds.

\paragraph*{T1 Multiple choice} The 209 items each have four options and one correct answer. Each distractor encodes a documented practitioner misconception, for example expecting more traces to rescue an attack whose leakage model is wrong, so option surface features give little help in guessing. T1 measures recall and applied pattern recognition.

\paragraph*{T2 Judgment} The 67 items each present a technical claim, which the model labels True or False and then justifies. For every claim the item author wrote four points that a sound justification must cover, so the item asks the model to explain why the verdict holds as well as to reach it.

\paragraph*{T3 Scenario} The 63 items each describe a realistic engineering situation with concrete parameters such as platform, sampling rate, trace count, and observed correlation values, and ask the model for an open-ended diagnosis. Each item carries an expert-written rubric of 3 to 5 dimensions with observable criteria, set by what its scenario calls for.

\paragraph*{T4 Code} The 41 tasks each give a function signature, a problem statement, and process constraints that rule out shortcuts, such as calling a library key-derivation function instead of implementing it or branching on a secret value. The model writes Python in 31 host-side tasks and C in 10 embedded tasks.

\subsection{Device-Condition Labels}\label{sec:labels}

Besides its sub-domain and task type, each item carries a label for how strongly its answer depends on the conditions of an IoT device. Such devices are often constrained nodes with tight limits on power, memory, and processing resources~\cite{rfc7228}, so a RAM budget or the setup of a leakage measurement can change what a correct answer must contain. The label has three levels.

L0 (general) items need no implementation-security knowledge. L1 (device-oriented) items need such knowledge, for example of side-channel analysis, countermeasures, or module certification, but no condition stated in the item decides the answer. L2 (device-constrained) items state a condition of the target device, such as its platform, a measurement, or a resource budget, whose removal changes what a correct answer must contain.

Table~\ref{tab:labels} gives the number of items at each level in each sub-domain.

\begin{table}[!t]
\centering
\caption{Device-Condition Labels by Sub-Domain}
\label{tab:labels}
\footnotesize
\begin{tabular}{@{}lrrrr@{}}
\toprule
Sub-domain & L0 & L1 & L2 & Total \\
\midrule
D1 Side-channel & 0 & 59 & 21 & 80 \\
D2 Fault injection & 0 & 35 & 16 & 51 \\
D3 Implementation & 32 & 29 & 22 & 83 \\
D4 Countermeasures & 1 & 41 & 20 & 62 \\
D5 Evaluation & 4 & 35 & 12 & 51 \\
D6 Integration & 21 & 18 & 14 & 53 \\
\midrule
Total & 58 & 217 & 105 & 380 \\
\bottomrule
\end{tabular}
\end{table}

\subsection{Item Authoring and Review}\label{sec:authoring}

Items are written by domain experts from the field's core textbooks~\cite{mangard2007power, joye2012fault}, the attack and countermeasure literature, and the certification standards, following a plan that fixes the sub-domain, sub-topic, and task type of every item. Each item sets a concrete situation in which knowledge must be applied, and states its units, context, and conditions precisely enough for experts to agree on the answer. Each item is then checked twice. Its writer re-derives the answer independently of the draft, and a second expert in side-channel analysis and cryptographic module evaluation reviews it. For code tasks, the reference implementation must also pass the test suite.

\subsection{Scoring}\label{sec:scoring}

\subsubsection{Per-Task Scoring} Multiple-choice items are scored by accuracy: an item is correct when the option letter in the model's answer matches the key. A judgment item is scored by the verdict gate, the rule that a wrong verdict scores the item zero however persuasive the justification reads, together with the justification rubric. The judge awards each of the four rubric dimensions 0, 0.5, or 1, and their mean is multiplied by the gate, which is 1 when the model's verdict matches the key and 0 otherwise. The dimension scores are kept even when the gate is 0, so that verdict accuracy and justification quality can be reported separately. A scenario item is scored by the mean of its three to five rubric dimensions, each awarded 0, 0.5, or 1 by the judge.

Submitted code is executed against the task's test suite, following execution-based grading in the HumanEval tradition~\cite{chen2021humaneval}. A code task is scored by task success, which is 1 when every functional test case passes and no process check fails, and 0 otherwise. A syntax error, a runtime error, or a timeout also scores 0. Process checks enforce the constraints stated in each task at the source level. For Python tasks they comprise an import allowlist, static scans for forbidden constructs such as branches or lookup tables, and checks that the required intermediate products are computed. For C tasks they add secret-taint rules and a static-memory budget, and the code must also compile for a Cortex-M0+ target.

The composite score $S$ of a model and its score $S_d$ on sub-domain $d$ are
\begin{equation}
S = \frac{1}{4}\sum_{t=1}^{4}\bar{s}_t, \qquad S_d = \frac{1}{4}\sum_{t=1}^{4}\bar{s}_{d,t},
\end{equation}
where $\bar{s}_t$ is the mean item score of the model on task type $t$ and $\bar{s}_{d,t}$ is the same mean over the items of sub-domain $d$ alone. The unweighted mean over task types keeps the 209 multiple-choice items from dominating the total.

\subsubsection{Judge Protocol} We score judgment and scenario answers with an LLM judge from a model family outside the evaluated set, which reduces the risk of self-enhancement bias, the tendency of a judge to favor answers from its own family~\cite{zheng2023judge}. For each response, the judge receives the claim or scenario, its rubric, and the response, but not the key of a judgment item or the reference answer of a scenario item. It returns a score for each rubric dimension, and all totals are computed in code. Each response is judged once, and this recorded pass gives the published score. Repeat passes, a second judge from another model family, and blind human re-scoring are used only to check the judge.

\section{Experiments}\label{sec:setup}

\subsection{Setup}

We evaluate 11 recent LLMs from nine vendors, nine open-weight and two proprietary. Table~\ref{tab:models} lists their vendor, weight availability, total and active parameter counts, and release date. We query the models through the SiliconFlow and OpenRouter APIs, each in its provider's default reasoning mode, with greedy decoding at temperature 0 and without tools. Each model answers every item from a zero-shot prompt, and one response per item is scored, which gives 380 responses per model and 4,180 in all. Python submissions run against their tests under pytest and Python 3.11, each in a separate working directory and within 120 seconds. C submissions are compiled with GCC 16.1 together with the hidden tests and a mock of the hardware interface, and the resulting test program must finish within 60 seconds. The Cortex-M0+ build uses arm-none-eabi-gcc 12.2. The judge is Qwen3.5-397B, and Grok-4.6 acts as the second judge in the reliability checks.

\begin{table}[H]
\vspace{-12pt}
\caption{Evaluated Models and Judges}
\label{tab:models}
\centering
\footnotesize
\setlength{\tabcolsep}{3pt}
\begin{tabular}{@{}llllr@{}}
\toprule
Model & Vendor & Weights & Params (active) & Released \\
\midrule
DeepSeek-V3 & DeepSeek & open & 671B (37B) & Dec 2024 \\
GLM-4-32B-0414 & Zhipu AI & open & 32B (32B) & Apr 2025 \\
Llama-4-Maverick & Meta & open & 400B (17B) & Apr 2025 \\
Hunyuan-A13B & Tencent & open & 80B (13B) & Jun 2025 \\
Ling-flash-2.0 & Ant Group & open & 100B (6.1B) & Sep 2025 \\
MiniMax-M2.5 & MiniMax & open & 230B (10B) & Feb 2026 \\
Kimi-K2.6 & Moonshot AI & open & 1T (32B) & Apr 2026 \\
DeepSeek-V4-Pro & DeepSeek & open & 1.6T (49B) & Apr 2026 \\
GLM-5.2 & Zhipu AI & open & 744B (40B) & Jun 2026 \\
GPT-5.6-Luna & OpenAI & closed & undisclosed & Jul 2026 \\
Gemini-3.7-Flash & Google & closed & undisclosed & Aug 2026 \\
\midrule
Qwen3.5-397B & Alibaba & open & 397B (17B) & Feb 2026 \\
Grok-4.6 & xAI & closed & undisclosed & Aug 2026 \\
\bottomrule
\end{tabular}
\end{table}
\subsection{Results and Analysis}\label{sec:analysis}

\begin{table*}[!t]
\caption{Scores in Percent by Task Type and by Sub-Domain}
\label{tab:overall}
\centering
\footnotesize
\setlength{\tabcolsep}{5pt}
\begin{tabular}{@{}lc@{\hspace{10pt}}cccc@{\hspace{10pt}}cccccc@{}}
\toprule
 & & \multicolumn{4}{c@{\hspace{10pt}}}{Task type} & \multicolumn{6}{c@{}}{Sub-domain} \\
\cmidrule(l{1pt}r{11pt}){3-6}\cmidrule(l{1pt}r{0pt}){7-12}
Model & Composite & T1 & T2 & T3 & T4 & D1 & D2 & D3 & D4 & D5 & D6 \\
\midrule
GLM-5.2 & \textbf{83.6} & \textbf{98.6} & \textbf{58.8} & 84.2 & 92.7 & 84.1 & \textbf{81.7} & \textbf{85.4} & 85.3 & 71.8 & \textbf{90.7} \\
Kimi-K2.6 & 82.5 & 97.1 & 58.0 & 82.2 & 92.7 & 83.5 & 80.6 & 75.6 & \textbf{85.5} & \textbf{80.5} & 89.7 \\
Gemini-3.7-Flash & 81.4 & \textbf{98.6} & 54.9 & 81.9 & 90.2 & \textbf{85.5} & 79.9 & 78.5 & 83.3 & 79.7 & 81.5 \\
GPT-5.6-Luna & 80.8 & 97.6 & 54.5 & \textbf{88.4} & 82.9 & 85.0 & 76.3 & 77.9 & 81.6 & 77.5 & 84.3 \\
DeepSeek-V4-Pro & 80.2 & 97.1 & 47.0 & 81.4 & \textbf{95.1} & 78.9 & 78.6 & 78.5 & 79.5 & 77.3 & 87.6 \\
MiniMax-M2.5 & 73.4 & 95.2 & 48.3 & 71.9 & 78.0 & 74.2 & 72.6 & 69.9 & 67.4 & 73.6 & 86.5 \\
DeepSeek-V3 & 68.4 & 91.9 & 41.2 & 67.5 & 73.2 & 63.5 & 66.1 & 62.5 & 73.2 & 68.3 & 79.3 \\
Ling-flash-2.0 & 61.3 & 83.3 & 40.1 & 55.9 & 65.9 & 61.6 & 61.5 & 61.4 & 53.8 & 58.8 & 73.5 \\
GLM-4-32B-0414 & 60.0 & 88.5 & 38.6 & 59.1 & 53.7 & 58.2 & 61.5 & 57.9 & 57.9 & 59.4 & 68.9 \\
Llama-4-Maverick & 59.7 & 93.3 & 34.9 & 57.1 & 53.7 & 57.3 & 67.1 & 49.6 & 56.9 & 60.2 & 73.8 \\
Hunyuan-A13B & 54.4 & 85.6 & 43.3 & 47.2 & 41.5 & 47.1 & 59.7 & 54.8 & 52.5 & 53.4 & 64.9 \\
\midrule
All 11 & 71.4 & 93.3 & 47.2 & 70.6 & 74.5 & 70.8 & 71.4 & 68.4 & 70.6 & 69.1 & 80.1 \\
\bottomrule
\end{tabular}
\end{table*}

\begin{figure*}[!t]
\centering
\includegraphics[width=\textwidth]{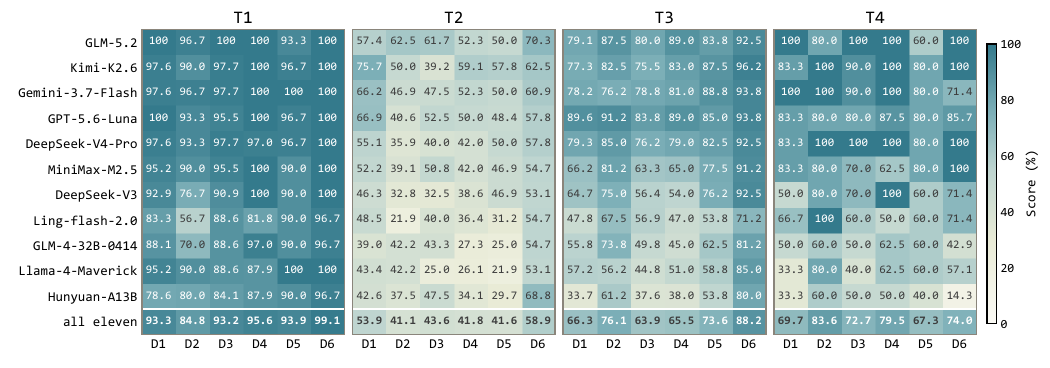}
\caption{Scores by sub-domain and task type for every model.}
\label{fig:hmap}
\end{figure*}

\subsubsection{Overview}\label{sec:results} Table~\ref{tab:overall} reports the scores of all 11 models by task type and by sub-domain, and Fig.~\ref{fig:hmap} resolves each score into cells of task type and sub-domain. GLM-5.2, Kimi-K2.6, Gemini-3.7-Flash, GPT-5.6-Luna, and DeepSeek-V4-Pro form a leading group with composite scores between 80.2\% and 83.6\%. The spread within this group is 3.4 points, half the 6.8-point gap between its lowest member and MiniMax-M2.5, which ranks sixth at 73.4\%. No pairwise difference within the group is statistically significant, whereas every member scores significantly higher than MiniMax-M2.5. The group comprises both proprietary models and three open-weight models. All six models released in 2026 rank above the five released earlier, whose composites range from 54.4\% to 68.4\%.

Scores differ far more across task types than across sub-domains. Pooled over the 11 models, the task scores range from 47.2\% on judgment to 93.3\% on multiple choice, whereas five of the six sub-domains lie between 68.4\% and 71.4\% and only D6 stands out at 80.1\%. The task types also spread the models apart to different degrees. The standard deviation of the 11 model scores is 5.1 points on multiple choice, 7.9 on judgment, 13.4 on scenario diagnosis, and 17.7 on code.

\subsubsection{Task Types}\label{sec:tasktypes}

\paragraph*{T1 Multiple choice} Multiple choice tests whether the model knows the fact, and every model scores highest on it. The five leading models reach 97.1\% to 98.6\%, and all 11 models answer 146 of the 209 items correctly. The remaining errors concentrate on a few items. Eight items are missed by five or more models, and on four of them seven models choose the same wrong option, which points to a misconception the models share. Multiple choice therefore still separates the weaker models, whose scores fall to 83.3\%, but no longer separates the leading ones.

\begin{figure}[!h]
\centering
\includegraphics[width=0.98\columnwidth]{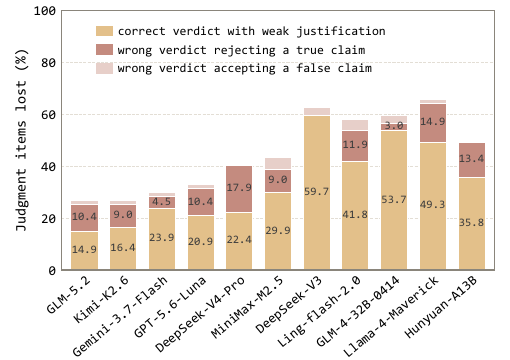}
\caption{Judgment items lost to a wrong verdict or a weak justification, per model.}
\label{fig:t2layers}
\end{figure}

\paragraph*{T2 Judgment} Judgment tests whether the model can decide if a security claim holds and explain why, and 10 of the 11 models score lowest on it. The models reach a correct verdict on 82.1\% to 97.0\% of the items, 88.5\% when pooled, well above the 62.7\% that always answering False would reach. The justifications behind these correct verdicts are much weaker. Their mean score is 53.4\% pooled, 57.3\% to 66.7\% for the five leading models and 41.1\% to 55.8\% for the other six, and only 6.0\% of them cover all four rubric points. Fig.~\ref{fig:t2layers} shows, for each model, the share of items lost to a wrong verdict or to a correct verdict whose justification falls below half marks. Verdict accuracy says little about justification quality. DeepSeek-V3 has the highest verdict accuracy, 97.0\%, yet the third-lowest justification score behind its correct verdicts, 42.5\%, and across the 11 models we find no association between the two measures, with a Spearman correlation of 0.04. The gated score, by contrast, correlates with the mean of the other three task scores at 0.81, whereas verdict accuracy correlates with it at 0.02.

\needspace{3\baselineskip}
The wrong verdicts lean in one direction. Of the 85 records without a correct verdict, 70 reject a true claim, 13 accept a false one, and 2 contain no verdict. Verdict accuracy is 74.5\% on true claims and 96.8\% on false ones, and 9 of the 11 models answer True less often than the 37.3\% of items whose key is True. One likely reason is that a claim counts as true only when every part of it holds, so a doubt about any part leads the model to reject the whole. The justifications behind wrong verdicts score between 0\% and 25.0\%.

Two same-family pairs show how a later model changes these scores. DeepSeek-V4-Pro scores 5.2 points higher than DeepSeek-V3 on multiple choice, 13.9 on scenario diagnosis, and 21.9 on code, and GLM-5.2 scores 10.1, 25.1, and 39.0 points higher than GLM-4-32B-0414. On judgment, verdict accuracy is lower in both later models, falling from 97.0\% to 82.1\% and from 94.0\% to 88.1\%. The justification behind correct verdicts rises from 42.5\% to 57.3\% and from 41.1\% to 66.7\%, and the gated score rises from 41.2\% to 47.0\% and from 38.6\% to 58.8\%. In both families the later model reaches fewer correct verdicts but justifies them better, a change that verdict accuracy alone would record as a decline.

\begin{table}[!h]
\vspace{-8pt}
\caption{Scores on Scenario Rubric Dimensions by Kind of Answer}
\label{tab:kinds}
\centering
\footnotesize
\setlength{\tabcolsep}{2.4pt}
\begin{tabular}{@{}lrccccc@{}}
\toprule
 & & \multicolumn{3}{c}{Mean score (\%)} & \multicolumn{2}{c}{Share (\%)} \\
\cmidrule(l{1pt}r{4pt}){3-5}\cmidrule(l{1pt}r{0pt}){6-7}
Kind & $n$ & Leading five & Other six & All & Fully met & Not met \\
\midrule
Mechanism & 83 & 87.3 & 68.7 & 77.2 & 65.4 & 11.1 \\
Assessment & 24 & 91.2 & 62.2 & 75.4 & 59.5 & 8.7 \\
Design & 83 & 83.6 & 55.7 & 68.4 & 55.6 & 18.8 \\
Standard & 28 & 79.3 & 57.1 & 67.2 & 51.9 & 17.5 \\
Diagnosis & 53 & 75.1 & 54.6 & 63.9 & 50.6 & 22.8 \\
Estimate & 11 & 74.5 & 49.2 & 60.7 & 43.8 & 22.3 \\
\bottomrule
\end{tabular}
\end{table}

\paragraph*{T3 Scenario} Scenario diagnosis tests whether the model can diagnose a concrete engineering situation. Scores range from 47.2\% to 88.4\%, and the highest belongs to GPT-5.6-Luna, which ranks fourth by composite score. The models largely agree on which scenarios are hard. The Spearman correlation between the item scores of two models averages 0.55 on scenario items, against 0.30 to 0.46 on the other task types, and no scenario receives full marks from all 11 models or zero from all of them.

To see what the models miss, we assigned each of the 282 rubric dimensions after the evaluation, from its rubric text, to one of six kinds of answer: explaining a mechanism, giving an overall assessment, proposing a design, stating what a standard requires, diagnosing a likely cause with a way to check it, or making a quantitative estimate. Each scenario uses only the kinds its task calls for. Table~\ref{tab:kinds} reports the mean score on each kind for the five leading models, the other six, and all 11, with the pooled shares of dimensions fully met and not met. Both groups show the same profile at different levels. The five leading models score 87.3\% on mechanism and 91.2\% on assessment dimensions but 74.5\% to 83.6\% on the other four kinds, and the other six score 68.7\% and 62.2\% against 49.2\% to 57.1\%. Pooled over the 11 models, a mechanism dimension is fully met in 65.4\% of cases, a diagnosis dimension in 50.6\%, and an estimate dimension in 43.8\%. The models are thus more reliable at explaining why something happens than at producing the specific design, requirement, check, or figure that an engineering answer needs.

\begin{figure}[!h]
\centering
\includegraphics[width=0.98\columnwidth]{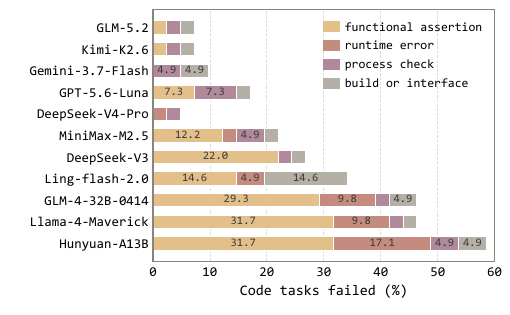}
\caption{Failed code tasks per model by grader class.}
\label{fig:t4stack}
\end{figure}

\paragraph*{T4 Code} Code tasks test whether the model can write an implementation that runs correctly. A task counts as solved only when all its functional tests pass and no process check fails, and the models solve 336 of the 451 model-task records. Fig.~\ref{fig:t4stack} shows the share of the 41 tasks each model fails, by the class the grader recorded. The 115 failures comprise 63 failed functional assertions, 19 runtime errors, 18 build or interface errors, and 15 process-check rejections, which flag shortcuts such as calling a prohibited primitive instead of implementing it. The five leading models fail 19 times in all, most often on process checks, whereas the other six fail 96 times, mostly on functional assertions and runtime errors, and Hunyuan-A13B alone accounts for seven of the runtime errors. The failed submissions are rarely near misses. Their functional pass rate has a median of 38\%, and 30.4\% of them pass no functional test. All 11 models solve 12 of the 41 tasks, and 13 of the 19 runtime errors fall on the six side-channel analysis tasks, which process trace arrays. The pooled success rate is 80.9\% on the 31 Python tasks and 54.5\% on the 10 C tasks.

\subsubsection{Sub-Domains}\label{sec:subdom} Pooled over the 11 models, D6 scores highest on multiple choice, judgment, and scenario items, as the bottom rows of Fig.~\ref{fig:hmap} show, and D2 scores highest on code. D6 covers key management, trusted execution, and public-key infrastructure, the part of the benchmark that draws least on the physical-attack literature. Among the other five sub-domains, the weakest one changes with the task type. The lowest pooled cell is D2 on multiple choice at 84.8\% and on judgment at 41.1\%, D3 on scenario diagnosis at 63.9\%, and D5 on code at 67.3\%. The per-model profiles in Table~\ref{tab:overall} vary in the same way. Nine of the 11 models score highest on D6, while the weakest sub-domain is D3 for five models, D4 and D5 for two each, and D1 and D2 for one each. GLM-5.2, which has the highest composite, scores 71.8\% on D5, below five other models.

\begin{table}[!t]
\caption{Items and Pooled Scores by Device-Condition Level}
\label{tab:iot}
\centering
\footnotesize
\setlength{\tabcolsep}{2.4pt}
\begin{tabular}{@{}lrcrcrcrc@{}}
\toprule
 & \multicolumn{2}{c}{T1} & \multicolumn{2}{c}{T2} & \multicolumn{2}{c}{T3} & \multicolumn{2}{c}{T4} \\
\cmidrule(l{3pt}r{1pt}){2-3}\cmidrule(l{2pt}r{2pt}){4-5}\cmidrule(l{2pt}r{2pt}){6-7}\cmidrule(l{2pt}r{1pt}){8-9}
Level & \multicolumn{1}{c}{$n$} & Score & \multicolumn{1}{c}{$n$} & Score & \multicolumn{1}{c}{$n$} & Score & \multicolumn{1}{c}{$n$} & Score \\
\midrule
L0 general & 33 & 93.4 & 8 & 65.1 & 8 & 80.2 & 9 & 88.9 \\
L1 device-oriented & 129 & 93.9 & 44 & 46.5 & 22 & 73.1 & 22 & 77.7 \\
L2 device-constrained & 47 & 91.9 & 15 & 39.9 & 33 & 66.6 & 10 & 54.5 \\
\midrule
All items & 209 & 93.3 & 67 & 47.2 & 63 & 70.6 & 41 & 74.5 \\
\bottomrule
\end{tabular}
\end{table}

\subsubsection{Device-Condition Levels}\label{sec:iot} Table~\ref{tab:iot} breaks the scores down by the device-condition level of each item. Two experts assigned the levels independently from the item text and its reference answer or rubric alone, agreeing on 90.3\% of the items with a Cohen's $\kappa$ of 0.83, with disagreements resolved by a third expert. The levels are spread unevenly across task types, with L2 covering 33 of the 63 scenario items but 47 of the 209 multiple-choice items, so we compare levels only within a task type.

Multiple choice scores alike at all three levels, 93.4\% on L0, 93.9\% on L1, and 91.9\% on L2, and none of the pairwise differences is significant. Code shows the largest gap. The L2 tasks score 54.5\%, against 77.7\% on L1 and 88.9\% on L0, and both differences are significant. Judgment and scenario diagnosis also score lowest on L2, at 39.9\% and 66.6\%, each 6.6 points below L1, but neither difference is significant.

The flat profile on multiple choice argues against a uniform difficulty shift across levels. On code, the L2 tasks are also the embedded C tasks, so the gap there reflects the language and the kind of task together with the device condition. The ranking of the models holds on the device-constrained items. Their order on the 33 L2 scenario items matches their order on all scenario items at a Spearman correlation of 0.99, and their composite over the 105 L2 items correlates with the full composite at 0.91.

\subsection{Scoring Reliability and Uncertainty}\label{sec:validation}

\subsubsection{Judge Reliability}\label{sec:judge} We check the judge against its recorded pass in three ways. A repeat pass and a second judge, Grok-4.6, re-score 44 judgment and 56 scenario responses from GLM-5.2, Kimi-K2.6, DeepSeek-V4-Pro, and MiniMax-M2.5. The first author re-scores 77 judgment and 174 scenario responses from all 11 models, the last batch drawn evenly across three bands of the judge's score. The rater was blind to the judge's scores and to the model names. Table~\ref{tab:judge} compares each check with the recorded pass on the normalized score of each response, with the bias taken as the recorded score minus the score of the check, in points.

\begin{table}[!h]
\vspace{-8pt}
\caption{Judge Reliability}
\label{tab:judge}
\centering
\footnotesize
\setlength{\tabcolsep}{5pt}
\begin{tabular}{@{}llrccc@{}}
\toprule
Check & Task & $n$ & ICC(2,1) & $\rho$ & Bias \\
\midrule
\multirow{2}{*}{Repeat pass} & T2 & 44 & 0.99 & 0.98 & $-$0.1 \\
 & T3 & 56 & 0.97 & 0.97 & $+$0.5 \\
\addlinespace[2pt]
\multirow{2}{*}{Second judge} & T2 & 44 & 0.95 & 0.93 & $+$1.3 \\
 & T3 & 56 & 0.83 & 0.91 & $+$2.3 \\
\addlinespace[2pt]
\multirow{2}{*}{Human rater} & T2 & 77 & 0.87 & 0.86 & $-$2.3 \\
 & T3 & 174 & 0.78 & 0.77 & $+$3.6 \\
\bottomrule
\end{tabular}
\end{table}

A repeat pass reproduces the recorded scores almost exactly. Grok-4.6 agrees closely on judgment and less closely on scenario responses, where it scores 2.3 points lower. The human rater agrees with the judge at an ICC of 0.87 on judgment and 0.78 on scenario responses. The judge scores 2.3 points below the human on judgment and 3.6 points above on scenarios, and only the scenario difference is significant. Across the 11 models, this difference shows no trend with the model's score, with Spearman correlations of $-0.28$ on judgment and $-0.08$ on scenarios.

\subsubsection{Uncertainty of the Results} Resampling the items within each task type 10,000 times places the 95\% interval of each composite about 3 to 5 points on either side of it. The 3.4-point spread within the leading group lies inside this range, so the order within the group is uncertain, and GLM-5.2 ranks first in 69\% of the resamples. Each of the five nonetheless scores significantly above MiniMax-M2.5. Of the 55 pairs of models, 37 still differ significantly after Holm correction for the number of comparisons, and MiniMax-M2.5, DeepSeek-V3, and Hunyuan-A13B keep their places in 95\% of the resamples.

The results hold under two other scoring choices. Weighting every item equally raises each composite by 5 to 13 points and keeps the five leading models in the same order. Scoring judgment by the justification alone, or with half credit for wrong verdicts, swaps only two adjacent pairs of models.

\section{Discussion}\label{sec:discussion}

\subsection{What the Justification Gap Means}

The central finding is that the models reach correct security verdicts far more often than they justify them. Pooled over the 11 models, 88.5\% of the verdicts are correct, but the justifications behind them score 53.4\%, and verdict accuracy shows no association with justification quality. In cryptographic engineering a recommendation that is right for the wrong reasons can pass review, be copied into the next design, and fail once the leakage model, the masking order, or the fault model changes. A benchmark that scores the verdict alone would record such an answer as correct, whereas the verdict gate keeps its weakness visible in the score.

Multiple choice has reached its ceiling for the strongest models. The five leading models score between 97.1\% and 98.6\%, so the format no longer separates them, whereas their scores spread over 11.8 points on judgment and 12.2 points on code. A domain benchmark that aims to track further progress therefore needs graded open-ended and executable tasks, together with a judging protocol whose reliability can be checked.

\subsection{Implications for IoT Security Practice}

We read the results as three working hypotheses for how LLMs can support the engineering of IoT device security. First, the leading models can draft well-specified implementations under expert review. They solve 87.1\% to 96.8\% of the Python tasks but 70\% to 90\% of the embedded C tasks, and their failed submissions most often break a process constraint that the task states. Second, their security verdicts need review before a design is signed off. Among the five leading models, 16.9\% to 27.3\% of the correct verdicts rest on a justification below half marks, and a correct verdict is the kind of answer a reviewer is least likely to question. Third, they can support security evaluation best by explaining mechanisms. The five leading models score 87.3\% on mechanism dimensions but 74.5\% to 79.3\% on estimates, diagnoses with a check, and the requirements of a standard, which are the parts of an evaluation that need verification.

\subsection{Limitations}

The results carry four limitations, concerning the judge, the API gateways, the distance from hardware, and the process checks. Each response is scored once by one judge, and the reliability checks rest on samples and a single human rater. Absolute judgment and scenario scores therefore depend on the judge, which scores scenario responses 3.6 points above the human rater, and comparisons among models are the firmer reading. The models and the judge ran through public API gateways, whose deployments and settings can change over an evaluation window. Every score is stored with its response, request configuration, and judge output, so it can be traced and re-checked. Device conditions are stated in the item text, and the embedded tasks check constraints at the source level, so the benchmark measures reasoning about a device rather than behavior on hardware. On the current code tasks the device condition and the C language coincide, and future tasks can separate the two. The process checks catch the shortcuts we anticipated, and a submission written to evade them could pass.

\subsection{Dual-Use Considerations}

Releasing attack-side items and analysis code raises the question of whether CESBench helps adversaries. We judge the added uplift low. Every attack item tests knowledge from the cited literature and standards, the Python tasks are analysis routines of the kind found in open-source side-channel toolkits, and the C tasks implement cryptographic and firmware security functions for an embedded target. What the benchmark adds is a measurement of how well models already handle this material, which defenders and evaluation laboratories need more than attackers do.

\section{Conclusion}\label{sec:conclusion}

CESBench is a benchmark of 380 expert-written items on the cryptographic engineering security of IoT devices, spanning six sub-domains and four task types, with judge-scored answers checked by repeat judging, a second judge, and blind human re-scoring. Across 11 open-weight and proprietary LLMs, composite scores range from 54.4\% to 83.6\%. The models label security claims correctly 88.5\% of the time, yet the justifications behind these correct verdicts score 53.4\%, and multiple choice no longer separates the five leading models while judgment and code still do. For IoT security practice, current LLMs can draft well-specified implementations and support evaluations under expert review, and their security verdicts need review before a design is signed off. Future work will add code tasks that separate device conditions from the programming language, a human baseline, and an error taxonomy over the archived responses.

\end{document}